\documentclass[prd,a4paper,preprint,preprintnumbers,nofootinbib,superscriptaddress]{revtex4-2}

\usepackage[a4paper, hdivide={1.91cm,,1.165cm}, vdivide={1.83cm,,3.0cm}]{geometry}
\usepackage[normalem]{ulem}

\usepackage[hyperfootnotes=false]{hyperref}
\hypersetup{linkcolor=red,
citecolor=green,
filecolor=cyan,
urlcolor=magenta}

\usepackage{amsfonts}

\usepackage{amsmath}
\usepackage{amssymb}

\usepackage{bbm}

\usepackage{bbold}

\usepackage{adjustbox}
\usepackage{booktabs}

\usepackage{colortbl}

\usepackage{cancel}

\usepackage{color}
\usepackage{xcolor}

\usepackage{graphicx}
\usepackage{xspace}

\usepackage{units}

\usepackage{slashed}

\usepackage{tensor}

\usepackage{gensymb}

\usepackage{cleveref}

\usepackage{tabularx}
\usepackage{multirow}

\usepackage{setspace}
\usepackage{orcidlink}

\usepackage{caption}
\usepackage{subcaption}

\usepackage{enumerate}
\usepackage{enumitem}

\usepackage[version=3]{mhchem}

\usepackage{lipsum}

\newcommand{{\ia} }{{\i}}
\newcommand{{\Ia} }{{\.I}}

\def\s2tw{{\rm sin ^2 \theta_{W}}}

\def\beq{\begin{equation}}
\def\eeq{\end{equation}}
\def\bea{\begin{eqnarray}}
\def\eea{\end{eqnarray}}

\def\bea{\begin{eqnarray}}
\def\eea{\end{eqnarray}}
\def\beeq{\begin{eqnarray}}
\def\eeeq{\end{eqnarray}}

\def\ba{\begin{array}}
\def\ea{\end{array}}

\def\xis0{{\Xi^{*0}}}

\def\g5{\gamma_5}

\allowdisplaybreaks[1]

\begin{document}


\title{$\Xi_c \to \Xi$ Semileptonic Decays: An LCSR View on the Experiment-Lattice Tension}

\author{T.~M.~Aliev\,\orcidlink{0000-0001-8400-7370}}
\email{taliev@metu.edu.tr}
\affiliation{Department of Physics, Middle East Technical University, Ankara, 06800, Turkey}

\author{S.~Bilmis\,\orcidlink{0000-0002-0830-8873}}
\email{sbilmis@metu.edu.tr}
\affiliation{Department of Physics, Middle East Technical University, Ankara, 06800, Turkey}
\affiliation{TUBITAK ULAKBIM, Ankara, 06510, Turkey}

\author{M.~Savci\,\orcidlink{0000-0002-6221-4595}}
\email{savci@metu.edu.tr}
\affiliation{Department of Physics, Middle East Technical University, Ankara, 06800, Turkey}

\date{\today}

\begin{abstract}  
 We present a light-cone QCD sum rule analysis of the semileptonic decays of 
$\Xi_c$ baryons, focusing on the channels $\Xi_c^0 \to \Xi^- \ell^+ \nu_\ell$, 
and  $\Xi_c^+ \to \Xi^0 \ell^+ \nu_\ell$. The transition form factors are 
calculated within the light-cone QCD sum rules framework, using the 
distribution amplitudes of the heavy $\Xi_c$ baryons. The obtained form 
factors are then used to compute the differential and total decay widths, 
as well as the branching fractions. Our numerical results for the branching 
fractions are $\mathcal{B}(\Xi_c^0 \to \Xi^- \ell^+ \nu_\ell) = (3.73 \pm 1.04)~\%$ ,
$\mathcal{B}(\Xi_c^0 \to \Xi^- \mu^+ \nu_\mu) = (3.59 \pm 1.01)~\%$,  
$\mathcal{B}(\Xi_c^+ \to \Xi^0 \ell^+ \nu_\ell) = (11.2 \pm 3.25)~\%$, and 
$\mathcal{B}(\Xi_c^+ \to \Xi^0 \mu^+ \nu_\mu) = (10.8 \pm 3.13)~\%$. 
These results are in good agreement with recent lattice QCD calculations, 
while being larger than the current experimental measurements and differing 
from the predictions of other theoretical approaches.

\end{abstract}



\maketitle


\newpage




\section{Introduction}
\label{sec:intro}


The semileptonic decays of heavy baryons are an important probe of heavy flavor
dynamics and the weak interaction. These processes involve a heavy-to-light quark
transition accompanied by a lepton–neutrino pair, and they provide valuable 
information on the underlying Cabibbo–Kobayashi–Maskawa (CKM) matrix elements 
as well as the structure of the effective weak Hamiltonian.  For example, 
the decays $\Xi_c \to \Xi \ell^+ \nu_\ell$ are governed at the quark level by the 
$c \to s$ transition and are sensitive to the CKM element $V_{cs}$. These modes 
also serve as a testing ground for heavy-quark symmetry in the baryon sector. 
Since the leptonic current in semileptonic decays is well understood theoretically, 
the primary uncertainties reside in the hadronic transition matrix elements. 
Accurate knowledge of the transition form factors is therefore crucial for 
interpreting current and future measurements of heavy baryon semileptonic decays.

Experimentally, charmed anti-triplet baryons (such as $\Lambda_c^+$, $\Xi_c^+$, 
and $\Xi_c^0$) have recently become accessible to precision studies, and 
significant progress has been reported. The absolute branching fractions of $\Xi_c^0$ and $\Xi_c^+$ were not directly 
measured; instead, they were determined relative to reference channels such as
$\Xi_c^0 \to \Xi^- \pi^+$ and $\Xi_c^+ \to \Xi^- \pi^+ \pi^+$. 
Recently the Belle Collaboration reported the first measurement of the absolute 
branching fraction for
$\Xi_c^{0} \to \Xi^{-}\pi^{+}$,  as $(1.80 \pm 0.50 \pm 0.14)\%$~\cite{Belle:2018kzz}.

Despite these recent experimental efforts, a notable discrepancy has emerged 
concerning the branching fractions of $\Xi_c \to \Xi \ell^{+}\nu_{\ell}$ decays. 
Recent measurements by the Belle Collaboration reported branching fractions of 
$\mathcal{B}(\Xi_c^{0} \to \Xi^{-}e^{+}\nu_{e}) = (1.31 \pm 0.04 \pm 0.07 \pm 0.38)\%$ and 
$\mathcal{B}(\Xi_c^{0} \to \Xi^{-}\mu^{+}\nu_{\mu}) = (1.27 \pm 0.06 \pm 0.10 \pm 0.37)\%$
~\cite{Belle:2021crz}. 
The ALICE Collaboration's measurement for 
$\mathcal{B}(\Xi_c^{0} \to \Xi^{-}e^{+}\nu_{e})$ was $(2.48 \pm 0.25 \pm 0.40 \pm 0.72)\%$
~\cite{ALICE:2021bli,Farrell:2025gis}, 
while the Particle Data Group (PDG) reports an average of
$(1.05 \pm 0.20)\%$~\cite{ParticleDataGroup:2020ssz}. 
These experimental values are considerably lower than expectations based on 
heavy-quark symmetry and flavor symmetry~\cite{Farrell:2025gis}. With more high-luminosity data expected from facilities like BESIII, LHCb, and Belle II, 
increasingly precise determinations of such branching ratios and decay distributions will 
become available.

This tension also extends to theoretical predictions. Various models—including lattice 
QCD~\cite{Farrell:2025gis}, relativistic quark models~\cite{Ebert:2011kk}, and flavor 
SU(3) approaches~\cite{He:2021qnc,Geng:2019bfz},  generally predict higher branching 
fractions than the current experimental average. 
For instance, a recent lattice QCD calculation predicts  
$\mathcal{B}(\Xi_c^{0} \to \Xi^{-} e^{+} \nu_{e}) = 3.58(12)\%$~\cite{Farrell:2025gis}, 
which explicitly notes that this is ``much higher than the more recent experimental 
results'' but ``reasonably close to the expectation from approximate SU(3) flavor symmetry.''
These discrepancies highlight the need for further theoretical and experimental 
investigations to improve our understanding of the nonperturbative aspects of 
charmed baryon decays.

On the theoretical side, the main challenge lies in calculating the form factors, which 
encapsulate the nonperturbative QCD effects. 
Several approaches have been used to study semileptonic heavy baryon decays. 
These include symmetry-based treatments using SU(3)$_f$ flavor symmetry
~\cite{Geng:2017mxn,Geng:2018plk,He:2021qnc}, 
constituent quark model calculations~\cite{Faustov:2018ahb,
Perez-Marcial:1989sch,Ivanov:1996fj,Geng:2019bfz}, 
lattice QCD simulations~\cite{Farrell:2025gis,Zhang:2021oja,Briceno:2012wt}, 
and QCD sum rule techniques~\cite{Zhao:2021sje,Liu:2010bh,Azizi:2011mw,Aliev:2021wat}. 
Each framework comes with its own advantages and limitations. 
For instance, SU(3) flavor symmetry can relate different decay channels, but being an 
approximate symmetry, it inherently induces on the order of 10\% theoretical uncertainty 
in decay amplitudes. Lattice QCD provides first-principles computations but is 
computationally intensive and has only recently begun to tackle charmed baryon form factors. 
Light-cone QCD sum rules (LCSR) offer a complementary approach: by expanding a suitable 
correlator near the light-cone, one can express the hadronic form factors in terms of 
universal baryonic light-cone distribution amplitudes (DAs) and perturbatively calculable 
hard kernels. This technique has been successfully applied to various problems in hadron 
physics, particularly in heavy-to-light transitions. For example, LCSR has been used to analyze the semileptonic decays of
$\Xi_c$ baryons~\cite{Liu:2010bh,Azizi:2011mw,Aliev:2021wat}, where the sum rules were 
formulated using the light-cone DAs of the final-state light baryons.

In this work, we perform an independent calculation of the transition form factors for 
the semileptonic decays $\Xi_c^0 \to \Xi^- \ell^+ \nu_\ell$, and $\Xi_c^+ \to \Xi^0 \ell^+ \nu_\ell$ 
using the  light-cone QCD sum rule approach.
By using the distribution amplitudes of the  initial charmed baryon $\Xi_c$, this strategy 
benefits from heavy quark symmetry, where the charm quark acts as a static color source at 
leading order, allowing a more controlled description of the baryon’s light-quark structure. 
All six relevant form factors for each transition are computed and used to predict decay 
widths and branching ratios.

The paper is organized as follows. In Section~\ref{sec:2}, we derive the light-cone sum 
rules for the six transition form factors. 
Section~\ref{sec:3} presents the numerical analysis, including predictions for decay widths 
and branching ratios, and compares our findings with results from the literature. 
Section~\ref{sec:conclusion} summarizes our main findings and highlights their implications 
for current and future experimental efforts.

\section{Calculation of the Baryonic Form Factors for $\Xi_c \to \Xi$ transition}
\label{sec:2}
In this section, we derive the light-cone sum rules (LCSR) for the transition form 
factors that describe the semileptonic decays of $\Xi_c$ charmed baryons into
light baryons, namely, $\Xi_c^0 \to \Xi^- \ell^+ \nu_\ell$ 
and $\Xi_c^+ \to \Xi^0 \ell^+ \nu_\ell$. 

These semileptonic decays are induced by the $c \to s$ transition and the matrix element 
can be written as :
\begin{equation}
  M = \frac{G_F}{\sqrt{2}} V_{cs} \langle \Xi(p') | \bar{c} \, \gamma_\mu (1 - \gamma_5) s | \Xi_c(p) \rangle
  \bar{\ell} \, \gamma^\mu (1 - \gamma_5) \nu,
\label{eq:1}
\end{equation}
where $G_F$ is the Fermi constant, and $V_{cs}$ is the element of the CKM matrix.

The hadronic matrix element 
$\langle \Xi(p') | \bar{c} \, \gamma_\mu (1 - \gamma_5) s | \Xi_c(p) \rangle$
is parameterized in terms of six independent form factors as follows:
\begin{align}
\langle \Xi(p') | \bar{c} \, \gamma_\mu (1 - \gamma_5) s | \Xi_c(p) \rangle 
&= \bar{u}_\Xi(p') \bigg\{ 
    f_1 \gamma_\mu 
    - i \frac{\sigma_{\mu \nu} q^\nu}{m_{\Xi_c}} f_2 
    + \frac{f_3 q_\mu}{m_{\Xi_c}} \notag \\
&\quad 
    - g_1 \gamma_\mu \gamma_5 
    + i \frac{\sigma_{\mu \nu} q^\nu}{m_{\Xi_c}} \gamma_5 g_2 
    - \frac{g_3 q_\mu}{m_{\Xi_c}} \gamma_5 
\bigg\} u_{\Xi_c}(p).
\label{eq:2}
\end{align}
Here $q = p - p'$ is the momentum transfer, and $f_i(q^2)$ and $g_i(q^2)$ (for $i=1,2,3$) 
are the vector and axial-vector form factors, respectively.

Since the form factors belong to the nonperturbative sector of QCD, their calculation 
requires a nonperturbative approach. Among the available methods, QCD sum 
rules~\cite{Shifman:1978bx,Shifman:1978by} hold an exceptional place, as they are firmly 
grounded in the fundamental QCD Lagrangian and offer a systematic framework for studying 
hadronic properties. In the present work, we use the light-cone QCD sum rule (LCSR) 
approach to calculate these form factors.

The main object of this method is the correlation function, which involves the time-ordered 
product of two operators: the interpolating current for the final-state baryon and the weak 
transition current responsible for the $c \to s$ transition. 
This correlation function is evaluated between the vacuum and the baryon state (in our case, 
the heavy baryon $\Xi_c$):

\begin{equation}
\Pi_\mu(p, q) = i \int d^4x e^{ip^\prime x} \langle 0 | T \big\{ \eta_\Xi(x)  \bar{s}(0) 
\gamma_\mu (1 - \gamma_5) c(0) \big\} | \Xi_c(p) \rangle,
\label{eq:3}
\end{equation}
where $\eta_\Xi$ is the interpolating current corresponding to the light baryon $\Xi$. 
The explicit form of this interpolating current is given by~\cite{Chung:1981cc}:
\begin{equation}
\eta_{\Xi}(x) = 2  \epsilon^{abc} \sum_{\ell=1}^{2} (s^{a} A^{\ell} d^{b}) B^{\ell} s^{c},
\end{equation}
where \( a, b, c \) are color indices, and the matrices \( A^\ell \) and \( B^\ell \) correspond to:
\[
A^1 = \mathbb{1}, \quad B^1 = \gamma_5, \qquad
A^2 = C \gamma_5, \quad B^2 = \beta \mathbb{1},
\]
with \( \beta \) being an arbitrary parameter and \( C \) is the charge conjugation operator. The choice \( \beta = -1 \) 
corresponds to the Ioffe current.

To derive the sum rules for the relevant form factors, we evaluate the correlation 
function in two different kinematic regions:  
in terms of hadronic degrees of freedom on one side, and in terms of quark and gluon 
fields using the operator product expansion (OPE) on the other.  
Matching the two representations through the dispersion relation allows us to extract 
the desired sum rules.
We begin by computing the hadronic representation of the correlation function.  
This is achieved by inserting a complete set of intermediate hadronic states with the 
same quantum numbers as the interpolating current \( \eta_\Xi \), and isolating the 
contribution from the ground-state \( \Xi \)-baryon. The result is:

\begin{equation}
\Pi_\mu(p, q) = \frac{ \langle 0 | \eta_\Xi(p') | \Xi(p', s') \rangle 
\langle \Xi(p', s') | J_\mu | \Xi_c(p, s) \rangle }{m_\Xi^2 - p'^2} + \cdots,
\end{equation}

where the ellipsis denotes contributions from higher resonances and continuum states.  
Using the definition of the matrix element
\begin{equation}
\langle 0 | \eta_\Xi(p^\prime) | \Xi(p', s') \rangle = \lambda_\Xi u(p', s'),
\end{equation}
together with the decomposition of the transition matrix element given in Eq.~(2), and 
summing over the spins of the final-state baryon,  
we derive the hadronic representation of the correlation function:
\begin{align}
\Pi_\mu^{\text{had}} &= \frac{\lambda_\Xi}{m_\Xi^2 - p'^2} (\slashed{p}' + m_\Xi)
\bigg\{
f_1 \gamma_\mu 
- i \frac{\sigma_{\mu \nu} q^\nu}{m_{\Xi_c}} f_2 
+ \frac{f_3 q_\mu}{m_{\Xi_c}} \notag \\
&\quad
- \left[ 
g_1 \gamma_\mu 
- i \frac{\sigma_{\mu \nu} q^\nu}{m_{\Xi_c}} g_2 
+ \frac{g_3 q_\mu}{m_{\Xi_c}} 
\right] \gamma_5 
\bigg\} u(p, s) + \cdots \quad,
\label{eq:9a}
\end{align}
where $\lambda_\Xi$ is the residue of the $\Xi$ baryon.
In the heavy quark limit, it is convenient to express the momentum of the initial 
heavy baryon as \( p_\mu = m_{\Xi_c} v_\mu \),  
where \( v_\mu \) is the four-velocity of the heavy baryon. Additionally, the heavy 
baryon spinor satisfies the relation \( \slashed{v} u(v) = u(v) \).  

To proceed further, we decompose the correlation function \( \Pi_\mu \) into 
Lorentz-invariant amplitudes (invariant functions), each multiplying a distinct 
Dirac structure:
\begin{align}
\Pi_\mu &= \Pi_1 v_\mu 
+ \Pi_2 v_\mu \gamma_5 
+ \Pi_3 \gamma_\mu  
+ \Pi_4 \gamma_\mu \gamma_5 
+ \Pi_5 q_\mu  
+ \Pi_6 q_\mu  \gamma_5 
+ \cdots
\end{align}

Below, we present the QCD-side expression for the correlation function corresponding 
to the \(\Xi_c \to \Xi\) transition.  
The QCD representation is obtained from Eq.~(1) by applying Wick’s theorem, which 
allows us to contract quark fields into propagators.  
This leads to the following expression:

\begin{equation}
  \label{eq:44}
  \begin{split}
    \Pi_\mu &= 2 \epsilon^{abc} \sum_{\ell=1}^2 \int d^4 x e^{i p^\prime x} 
(A^\ell)_{\alpha \beta} (B^\ell)_{\rho \gamma} (\gamma_\mu (1-\gamma_5))_{h \phi} \\
    & \bigg \{ S_{\gamma h} \langle 0 | s_\alpha^a(x) d_{\beta}^b(x) c_\phi^c(0) | \Xi_c \rangle +
    S_{\alpha h} \langle 0 | s_\gamma^a(x) d_\beta^b(x) c_\phi^c(0) | \Xi_c \rangle
\bigg\},
  \end{split}
\end{equation}
where S is the strange quark propagator.
As seen from Eq.\eqref{eq:44}, the evaluation of the correlation function 
requires knowledge of the matrix element
\[
\epsilon ^{abc} \langle 0 | s^a_\alpha(x) d^b_\beta(x) c^c_\gamma(0) | \Xi_c(v) \rangle.
\]
which is expressed in terms of the light-cone distribution 
amplitudes (DAs) of the \(\Xi_c\) baryon  \cite{Ali:2012pn} in the following
way:
\begin{equation}
\epsilon^{abc} \langle 0 | s^a_\alpha(t_1 n) d^b_\beta(t_2 n) h^c_\gamma(0) | \Xi_c(v) \rangle 
= \sum_{j=1}^{4} \mathcal{A}_j (\Gamma_j)_{\alpha \beta} (u_j(v))_\gamma,
\label{eq:11a}
\end{equation}
where \( h^c_\gamma \) is the heavy quark field in HQET, \( \Gamma_j \) are Dirac 
structures, \( \mathcal{A}_j \) represents distribution amplitudes, 
and \( u_j(v) \) are spinors describing the light degrees of freedom in the heavy baryon.

We emphasize that these DAs are derived within the framework of heavy quark 
effective theory (HQET). The relation between the heavy baryon states in full 
QCD and those in HQET is given by:
\[
| \Xi_c(p) \rangle = \sqrt{m_{\Xi_c}} \, | \Xi_c(v) \rangle,
\]
where \( v \) is the four-velocity of the heavy baryon.

The Dirac structures \(\Gamma_j\) and their associated coefficients \(\mathcal{A}_j\) 
in Eq.~\eqref{eq:11a} are given as follows:

\begin{align}
\mathcal{A}_1 &= \frac{1}{8} f^{(2)} \psi_2(t_1, t_2), & \Gamma_1 &= \slashed{n} 
\gamma_5 C^{-1}, \notag \\
\mathcal{A}_2 &= -\frac{1}{8} f^{(1)} \psi_3^\sigma(t_1, t_2), & \Gamma_2 &= i 
\sigma_{\mu \nu} \bar{n}^\mu n^\nu \gamma_5 C^{-1}, \notag \\
\mathcal{A}_3 &= \frac{1}{4} f^{(1)} \psi_3^s(t_1, t_2), & \Gamma_3 &= \gamma_5 C^{-1}, \notag \\
\mathcal{A}_4 &= \frac{1}{8} f^{(2)} \psi_4(t_1, t_2), & \Gamma_4 &= \slashed{n} \gamma_5 C^{-1}~.
\end{align}

The functions \( \psi_2 \), \( \psi_3^s \), \( \psi_3^\sigma \), and \( \psi_4 \) are 
light-cone distribution amplitudes (DAs) of twist 2, 3, and 4, respectively.

The light-cone vectors $n_\mu$ and $\bar{n}_\mu$ are defined as:
\begin{align}
n_\mu &= \frac{x_\mu}{v  x},  \\
\bar{n}_\mu &= 2 v_\mu - n_\mu . 
\label{eq:13a}
\end{align}

The DAs \( \psi(t_1, t_2) \) are related to their momentum-space counterparts through a 
double Fourier transform:

\begin{equation}
\psi(t_1, t_2) = \int_0^\infty \omega d\omega \int_0^1 du\, e^{-i \omega (t_1 \bar{u} + 
t_2 u)} \psi(\omega, u),
\end{equation}

where \( \omega \) is the total light-cone momentum of the two light quarks, 
and \( \bar{u} = 1 - u \).  

In our case, \( t_1 = t_2 = v  x \), which simplifies the expression
as follows:
\begin{equation}
\psi(t_1, t_2) = \int_0^\infty \omega d\omega \int_0^1 du\, e^{-i \omega v x} \psi(\omega, u).
\end{equation}
The explicit expressions of these DAs \( \psi_2(\omega, u), \psi_3^s(\omega, u), 
\psi_3^\sigma(\omega, u), \psi_4(\omega, u) \)  
for the \( \Xi_c \)  baryon have been derived in  \cite{Ali:2012pn}, and serve as 
inputs in our calculation.

Combining Eqs.~\eqref{eq:44} and \eqref{eq:11a}, the QCD-side expression for the 
correlation function takes the form:

\begin{align}
\Pi_\mu^{\text{QCD}} 
&= 2i\,  
   \int d^4x \int_0^1 du \int_0^\infty \omega d\omega \,
   e^{i(p^\prime - \omega v) x} 
   \sum_{\ell = 1}^{2} \sum_{j = 1}^{4} \mathcal{A}_j  \nonumber \\[6pt]
&\quad \times 
   \Big\{ 
      \text{Tr}\!\left[\Gamma_j A^{\ell}  \right] B^\ell 
      S \gamma_\mu (1 - \gamma_5) + 
      \big[ B^\ell \Gamma_j^T A^{\ell\,T} S \gamma_\mu (1 - \gamma_5) \big]  
   \Big\} u(v).
\label{eq:QCDcorr}
\end{align}
After performing integration over \( x \), the correlation function is 
expressed in terms of the DAs and propagators in momentum space.

To extract the form factors, we match the coefficients of the relevant Lorentz 
structures appearing in both the hadronic and QCD representation of the correlation 
function. We choose following structures:

\[
v_\mu , v_\mu \gamma_5,  \gamma_\mu, \gamma_\mu \gamma_5, q_\mu ,\mbox{and }  q_\mu  \gamma_5~.
\]

Matching the coefficients of these Lorentz structures allows us to derive the sum 
rules for the transition form factors as:

\begin{align}
\frac{2\lambda m_{\Xi_c}}{m_{\Xi}^2 - p^{\prime 2}} f_{1} &= \Pi_{1}^{\text{QCD}} \nonumber \\[6pt]
- \frac{2\lambda m_{\Xi_c}}{m_{\Xi}^2 - p^{\prime 2}} g_{1} &= \Pi_{2}^{\text{QCD}} \nonumber \\[6pt]
\frac{\lambda}{m_{\Xi}^2 - p^{\prime 2}} \Big[ (m_{\Xi} - m_{\Xi_c}) \Big( f_{1} + 
\frac{f_{2}}{m_{\Xi_c}} (m_{\Xi} + m_{\Xi_c}) \Big) \Big] 
&= \Pi_{3}^{\text{QCD}} \nonumber \\[6pt]
- \frac{\lambda}{m_{\Xi}^2 - p^{\prime 2}} \Big[ (m_{\Xi} + m_{\Xi_c}) \Big(g_{1} + 
\frac{g_{2}}{m_{\Xi_c}} (m_{\Xi} - m_{\Xi_c})\Big) \Big] 
&= \Pi_{4}^{\text{QCD}} \nonumber \\[6pt]
\frac{\lambda}{m_{\Xi}^2 - p^{\prime 2}} \Big[ -2 f_{1} - \frac{f_{2}}{m_{\Xi_c}} (m_{\Xi} + m_{\Xi_c}) + 
\frac{f_{3}}{m_{\Xi_c}} (m_{\Xi} + m_{\Xi_c}) \Big] 
&= \Pi_{5}^{\text{QCD}} \nonumber \\[6pt]
\frac{\lambda}{m_{\Xi}^2 - p^{\prime 2}} \Big[ 2 g_{1} + \frac{g_{2}}{m_{\Xi_c}} (m_{\Xi} - m_{\Xi_c}) - 
\frac{g_{3}}{m_{\Xi_c}} (m_{\Xi} - m_{\Xi_c}) \Big] 
&= \Pi_{6}^{\text{QCD}} \label{eq:Pi}
\end{align}

The invariant amplitudes \( \Pi_i^{\text{QCD}} \) can be expressed in terms of 
dispersion integrals over the spectral densities as:
\begin{equation}
\Pi_i^{\text{QCD}} = \int_0^1 du \int_0^\infty d\sigma\, \sigma 
\left\{
\frac{\rho_i^{(1)}}{\bar{\sigma}\Delta} + \frac{\rho_i^{(2)}}{ \bar{\sigma}^2 \Delta^2} + 
\frac{ \rho_i^{(3)}}{\bar{\sigma}^3 \Delta^3}
\right\},
\end{equation}

with
\begin{equation}
\Delta = p'^2 - s(\omega), \quad \sigma = \frac{\omega}{m_\Xi}, \quad
s(\omega) = \frac{m_s^2 + m_\Xi^2 \bar{\sigma} \sigma - 
\sigma q^2}{\bar{\sigma}}, \quad \bar{\sigma} = 1 - \sigma.
\end{equation}
To suppress the contributions of higher resonances and the continuum, 
we apply a Borel transformation with respect to \( p'^2 \). 
Matching the hadronic and QCD sides of each invariant amplitude  yields 
the desired sum rules for the transition form factors.

The Borel transformation and continuum subtraction are performed using the 
following master formula~\cite{Gubernari:2018wyi,Aliev:2019ojc}:
\begin{align}
\int_0^{s_0} ds\, \frac{I_n}{(p'^2 - s)^n} 
&= \int_0^{\sigma_0} d\sigma\, (-1)^n \frac{e^{-s(\sigma)/M^2} I_n}{(n-1)! (M^2)^{n-1}} \notag \\
&\quad + \frac{(-1)^n}{(n-1)!} e^{-s/M^2} \sum_{j=1}^{n-1} \frac{1}{(M^2)^{n-j-1}} \frac{1}{s^\prime} 
\left( \frac{d}{d\sigma} \frac{1}{s^\prime} \right)^{j-1} I_n,
\end{align}

with
\[
s^\prime = \frac{ds}{d\sigma}, \quad 
\left( \frac{d}{d\sigma} \frac{1}{s^\prime} \right)^n I_n \Rightarrow 
\text{nested derivatives acting on } I_n  \text{ and } I_n = \frac{\sigma
\rho^{(n)}}{\bar{\sigma}^n}.
\]
Here, \( \sigma_0 \) is the solution of the equation \( s(\omega) = s_0 \), with \( s_0 \) 
denoting the continuum threshold. Since the Borel-transformed invariant functions \( \Pi_i \) 
have lengthy expressions, we do not present them explicitly here.

At the end of this section, we provide the expressions for the differential
decay width of the semileptonic transition.  
To this end, we employ the helicity amplitude formalism (see \cite{Gutsche:2015mxa}).

These amplitudes are conveniently calculated in the rest frame of the initial heavy baryon, 
with the $z$-axis aligned along the momentum of the off-shell $W$ boson.

The vector current helicity amplitudes are given as:

\begin{align}
H^{V}_{+\frac{1}{2}, t} &= \frac{\sqrt{Q_+}}{\sqrt{q^2}} \left( m_- f_1^V + 
\frac{q^2}{m_\Xi} f_3^V \right), \nonumber \\
H^{V}_{+\frac{1}{2}, +1} &= \sqrt{2 Q_-} \left( f_1^V + 
\frac{m_+}{m_\Xi} f_2^V \right), \nonumber \\
H^{V}_{+\frac{1}{2}, 0} &= \frac{\sqrt{Q_-}}{\sqrt{q^2}} 
\left( m_+ f_1^V + \frac{q^2}{m_\Xi} f_2^V \right)~,
\end{align}
where \( m_\pm = m_{\Xi_c} \pm m_{\Xi} \), and \( Q_\pm =  m_\pm^2 - q^2 \).

The axial-vector helicity amplitudes are obtained from the vector ones through 
the substitutions:
\begin{align}
H^{A}_{+\frac{1}{2}, t} &= H^{V}_{+\frac{1}{2}, t} \big|_{Q_+ \to Q_-,\, m_- \to m_+,\, f_i^V \to g_i^V,\, f_3^V \to -g_3^V}, \nonumber \\
H^{A}_{+\frac{1}{2}, +1} &= H^{V}_{+\frac{1}{2}, +1} \big|_{Q_- \to Q_+,\, m_+ \to m_-,\, f_1^V \to g_1^V,\, f_2^V \to -g_2^V}, \nonumber \\
H^{A}_{+\frac{1}{2}, 0} &= H^{V}_{+\frac{1}{2}, 0} \big|_{Q_- \to Q_+,\, m_+ \to m_-,\, f_1^V \to g_1^V,\, f_2^V \to -g_2^V},
\end{align}
From parity considerations, the helicity amplitudes satisfy the relation:
\[
H^{V(A)}_{-\lambda, -\lambda_W} = \pm H^{V(A)}_{\lambda, \lambda_W},
\]

where the first index refers to the helicity of the final-state (daughter) baryon, 
and the second to the $W$ boson.

\vspace{1em}
After standard calculations, the differential decay width takes the form:

\begin{equation}
\frac{d\Gamma}{dq^2} = \Gamma_0 \frac{(q^2 - m_\ell^2)^2}{m_{\Xi_c}^7 q^2} |\vec{p}|\, H_{\text{tot}},
\end{equation}

where the total helicity amplitude is:

\begin{equation}
H_{\text{tot}} = H_U + H_L + \frac{m_\ell^2}{2 q^2} \left( H_U + H_L + 3 H_S \right),
\end{equation}
where
\begin{align}
H_U &= \left| H_{+\frac{1}{2}, +1} \right|^2 + \left| H_{-\frac{1}{2}, -1} \right|^2, \notag \\
H_L &= \left| H_{+\frac{1}{2}, 0} \right|^2 + \left| H_{-\frac{1}{2}, 0} \right|^2, \notag \\
H_S &= \left| H_{+\frac{1}{2}, t} \right|^2 + \left| H_{-\frac{1}{2}, t} \right|^2.
\end{align}

The full helicity amplitude is defined as the difference between vector 
and axial-vector components:
\begin{equation}
H_{\lambda, \lambda_W} = H^V_{\lambda, \lambda_W} - H^A_{\lambda, \lambda_W},
\end{equation}

and the overall normalization factor \( \Gamma_0 \) is:
\begin{equation}
  \Gamma_0 = \frac{G_F^2 |V_{cs}|^2 m_{\Xi_c}^5}{192 \pi^3}.
  \label{eq:decayrate}
\end{equation}

The three-momentum magnitude of the final-state baryon is:
\[
|\vec{p}| = \frac{1}{2 m_{\Xi_c}} \lambda^{1/2}(m_{\Xi_c}^2, m_{\Xi}^2, q^2),
\]

where \( \lambda(a, b, c) \) is the usual Källén function:
\[
\lambda(a, b, c) = a^2 + b^2 + c^2 - 2ab - 2ac - 2bc.
\]
\section{Numerical Analysis}
\label{sec:3}

In this section, we present the numerical analysis of the light-cone sum rules 
for the transition form factors. Using the derived form factors, we estimate 
the branching ratios for the semileptonic decays $\Xi_c \to \Xi \ell \nu$. 
The central nonperturbative inputs in the LCSR framework are the distribution 
amplitudes (DAs) of the initial heavy baryon. As mentioned earlier, we employ 
the $\Xi_c$ baryon DAs derived in~\cite{Ali:2012pn}, whose explicit forms are given by:  
\begin{align}
\psi_2(u,\omega) &= \omega^2 u \bar{u} \sum_{i=0}^2 \frac{a_i}{\epsilon_i^4} 
\frac{C_i^{3/2}(2u-1)}{|C_i^{3/2}|^2} e^{-\omega/\epsilon_i}, \nonumber \\[6pt]
\psi_3^{(\sigma,s)}(u,\omega) &= \frac{\omega}{2} \sum_{i=0}^2 \frac{a_i}{\epsilon_i^3} 
\frac{C_i^{1/2}(2u-1)}{|C_i^{3/2}|^2} e^{-\omega/\epsilon_i}, \nonumber \\[6pt]
\psi_4(u,\omega) &= \sum_{i=0}^2 \frac{a_i}{\epsilon_i^2} 
\frac{C_i^{1/2}(2u-1)}{|C_i^{1/2}|^2} e^{-\omega/\epsilon_i},
\end{align}
where \( C_i^{\ell}(2u - 1) \) are Gegenbauer polynomials, and the normalization 
factors are defined as:
\[
|C_i^{\ell}|^2 = \int_0^1 du\, \left| C_i^{\ell}(2u - 1) \right|^2.
\]
The numerical values of the shape parameters \( a_i \) and \( \epsilon_i \) are taken 
from~\cite{Ali:2012pn}.  
The normalization constants \( f^{(1)} \) and \( f^{(2)} \) which appear in the 
definitions of \( a_i \), are given by~\cite{Ali:2012pn}:

\[
f^{(1)} = f^{(2)} = (2.23 \pm 0.35) \times 10^{-2}~\text{GeV}^3.
\]

Additional input parameters used in the numerical analysis are adopted from the 
Particle Data Group (PDG)~\cite{ParticleDataGroup:2020ssz} as summarized in Table~\ref{tab:inputvalues}.

\begin{table}[h]
\centering
\caption{Input parameters used in the numerical analysis (values 
from ~\cite{ParticleDataGroup:2020ssz}).}
\begin{tabular}{lc}
\toprule
$|V_{cs}|$ & $0.975 \pm 0.006$ \\
$m_c(\overline{m_c})$ & $1.273 \pm 0.046~\text{GeV}$ \\
$m_{\Xi^0_c}$ & $2470.44 \pm 0.28~\text{MeV}$ \\
$m_{\Xi^+_c}$ & $2467.71 \pm 0.23~\text{MeV}$ \\
$m_{\Xi^-}$ & $1321.71 \pm 0.07~\text{MeV}$ \\
$m_{\Xi^0}$ & $1314.86 \pm 0.20~\text{MeV}$ \\
$\tau_{\Xi_c^0}$ & $150.4 \pm 2.8 ~\text{fs}$ \\
$\tau_{\Xi_c^+}$ & $453 \pm 5~\text{fs}$ \\
\bottomrule
\end{tabular}
    \label{tab:inputvalues}
\end{table}

In addition to the above inputs, the sum rule expressions involve three auxiliary 
parameters:
\begin{itemize}
    \item the Borel mass parameter \( M^2 \),
    \item the continuum threshold \( s_0 \), and
    \item the mixing parameter \( \beta \), which appears in the interpolating current. 
In numerical calculations, we used the Ioffe current, i.e, $\beta= -1$.
\end{itemize}
A detailed discussion of the stability of the results with respect to the variation 
of these auxiliary parameters is provided below.

The continuum threshold \( s_0 \) is chosen such that the mass obtained from the 
two-point QCD sum rule reproduces the experimental baryon mass within 10\% accuracy. 
Our numerical analysis shows that this condition is satisfied when:

\[
 s_0 = (3.5 \pm 0.5)~\text{GeV}^2.
\]

The working region of the Borel parameter \( M^2 \) is determined by imposing that 
both the higher-twist corrections and the continuum contributions remain subdominant 
compared to the leading-twist term. This condition ensures that the dominant 
contribution to the sum rule arises from the lowest-twist DA.

Under these criteria, our numerical analysis yields the following working window 
for the Borel mass parameter:
\[
2.0~ \text{GeV}^2 \leq M^2 \leq 3.0~\text{GeV}^2.
\]

Once the working regions of the auxiliary parameters \( M^2 \) and \( s_0 \)  are 
established, we proceed to analyze the \( q^2 \)-dependence of the form factors.
Note that the LCSR predictions are valid only in the low-\( q^2 \) region and must 
be extrapolated to cover the full physical kinematic range:
\[
m_\ell^2 \leq q^2 \leq (m_{\Xi_c} - m_{\Xi})^2.
\]

In particular, the reliability of the sum rules deteriorates at higher \( q^2 \) values.  
The range where the LCSR calculation remains reliable is:

\[
q^2 \leq 0.5~\text{GeV}^2.
\]
To extrapolate the LCSR predictions to the entire kinematic range, we use the 
model-independent \( z \)-series expansion (Boyd-Grinstein-Lebed or 
BGL approach)~\cite{Bourrely:2008za}.

The conformal mapping is defined as:
\[
z(q^2) = \frac{\sqrt{t_+ - q^2} - \sqrt{t_+ - t_0}}
               {\sqrt{t_+ - q^2} + \sqrt{t_+ - t_0}},
\]
where
\[
t_+ = (m_{\Xi_c} + m_\Xi)^2, \qquad
t_0 = (m_{\Xi_c} - m_\Xi)^2.
\]

We find that the LCSR predictions for the form factors are best reproduced by 
the following fit function:
\[
f(q^2) = \frac{1}{1 - q^2/m_{\text{pole}}^2} 
\left\{ a_0^f + a_1^f\, z(q^2) + a_2^f\, z^2(q^2) \right\},
\]
where \( m_{\text{pole}} \) corresponds to the mass of the lowest-lying resonance with 
the same quantum numbers as the current involved in the transition. The pole masses 
used in the fits are:

\[
m_{\text{pole}} =
\begin{cases}
2.112~\text{GeV} & \text{for } f_1, f_2, \\
2.535~\text{GeV} & \text{for } g_1, g_2, \\
2.317~\text{GeV} & \text{for } f_3, \\
1.969~\text{GeV} & \text{for } g_3~.
\end{cases}
\]

The fit parameters \( a_0^f, a_1^f, a_2^f \) for each form factor are extracted 
by performing a least-squares match of the parameterization to the LCSR predictions 
in the region \( 0 \leq q^2 \leq 0.5~\text{GeV}^2 \). The resulting fits are employed 
to extend the form factors to the entire kinematic range, enabling reliable decay 
width and branching ratio predictions.
Table~\ref{tab:ffvalues} summarizes the form factor values at \( q^2 = 0 \) 
for the $\Xi_c \to \Xi$ transitions.

\begin{table}[htb]
  \centering
  \caption{Form factors $f_{i}$ and $g_i$ at $q^2 = 0$ for the 
$\Xi_c^0 \to \Xi^- \ell^+ \nu_\ell$ transition.}
  \begin{tabular}{cc}
    \toprule
    $f_1$ &  $0.84 \pm 0.11$ \\
    $f_2$ &  $0.49 \pm 0.06$ \\
    $f_3$ & $-0.35 \pm 0.11$ \\
    $g_1$ &  $0.84 \pm 0.11$ \\
    $g_2$ &  $0.49 \pm 0.06$ \\
    $g_3$ & $-0.42 \pm 0.08$ \\
    \bottomrule
  \end{tabular}
    \label{tab:ffvalues}
\end{table}

The uncertainties of the fit parameters are estimated through a Monte Carlo simulation. 
We generated 5000 pseudo-experiments by randomly sampling the input parameters within 
their uncertainties, and computed the corresponding form factors at \( q^2 = 0 \) 
in each case. As an illustration, Fig.~1 shows the distribution of the form factor values 
obtained from the ensemble. The resulting distributions were fit with Gaussians to extract 
central values and standard deviations, which are quoted as the uncertainties 
in Table~\ref{tab:ffvalues}. After determining the form factors, we compute the total 
decay widths and branching ratios using Eq.~\eqref{eq:decayrate}. The lifetimes of the 
charmed baryons required for these calculations are taken from
Table~\ref{tab:inputvalues}.

Based on the fitted form factors and baryons lifetimes, the branching ratios for the 
semileptonic decays are computed as follows:
\begin{align*}
\mathcal{B}(\Xi_c^0 \to \Xi^- e^+ \nu_e) &= (3.73 \pm 1.04) \%, \\
\mathcal{B}(\Xi_c^0 \to \Xi^- \mu^+ \nu_\mu) &= (3.59 \pm 1.01)~\%, \\
\mathcal{B}(\Xi_c^+ \to \Xi^0 e^+ \nu_e) &= (11.20 \pm 3.25) \%, \\
\mathcal{B}(\Xi_c^+ \to \Xi^0 \mu^+ \nu_\mu) &= (10.8 \pm 3.13) \%. \\
\end{align*}
We note that, in estimating the form factors and the corresponding branching ratios, we also calculated the form factors obtained from alternative Lorentz structures. Our numerical analysis shows that the resulting branching ratios are very close to each other; in particular, the differences are below 4\%

%
As has already been discussed, semileptonic $\Xi_c$ decays have been examined using various 
theoretical frameworks. For comparison, Table~\ref{tab:comparison} compiles the 
predicted branching ratios from these approaches along with available experimental 
measurements.
\begin{figure}[t]
\includegraphics[width=0.32\textwidth]{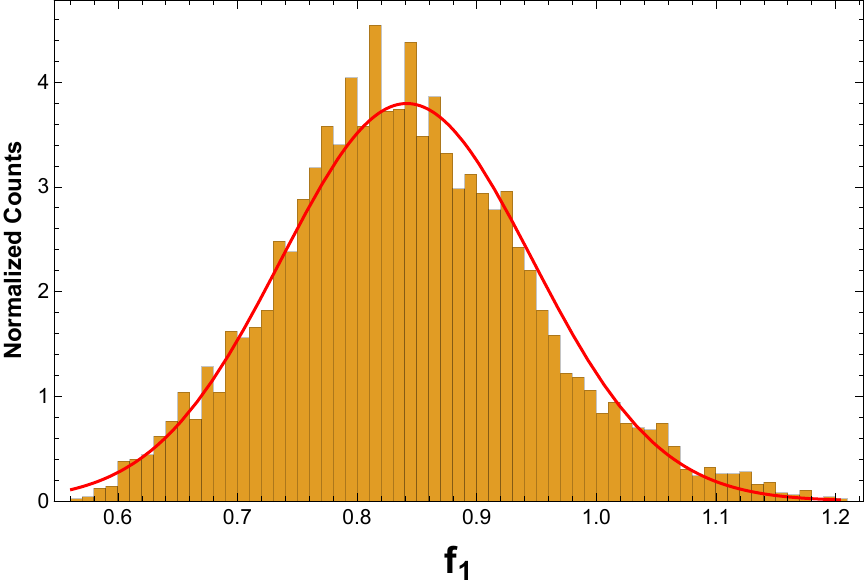}
\includegraphics[width=0.32\textwidth]{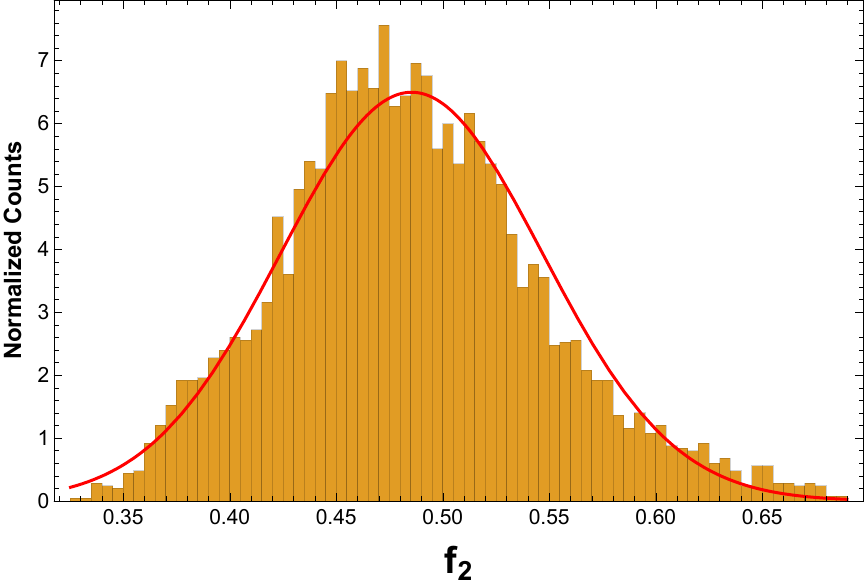}
\includegraphics[width=0.32\textwidth]{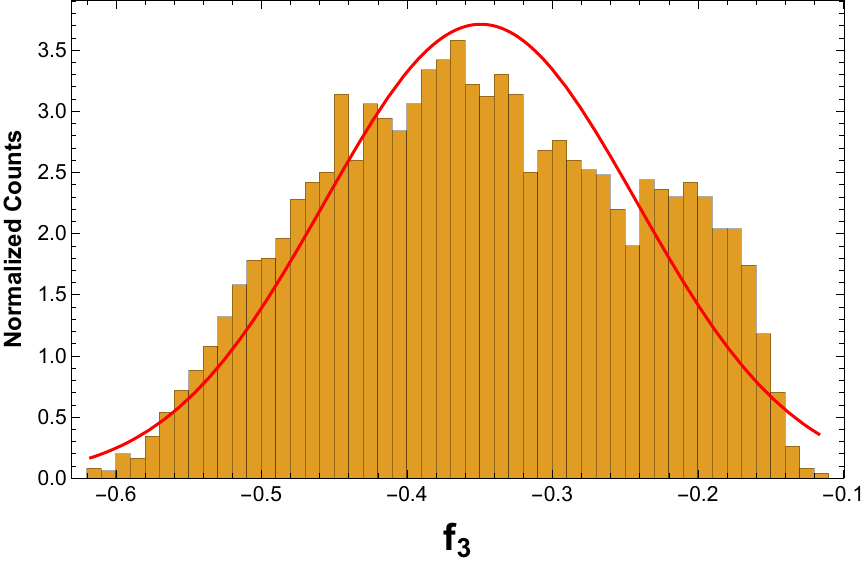} \\
\includegraphics[width=0.32\textwidth]{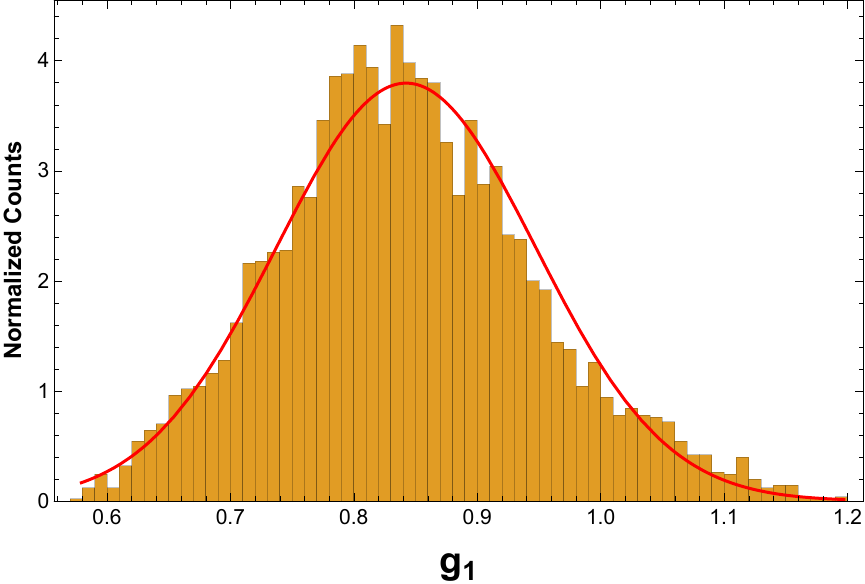} 
\includegraphics[width=0.32\textwidth]{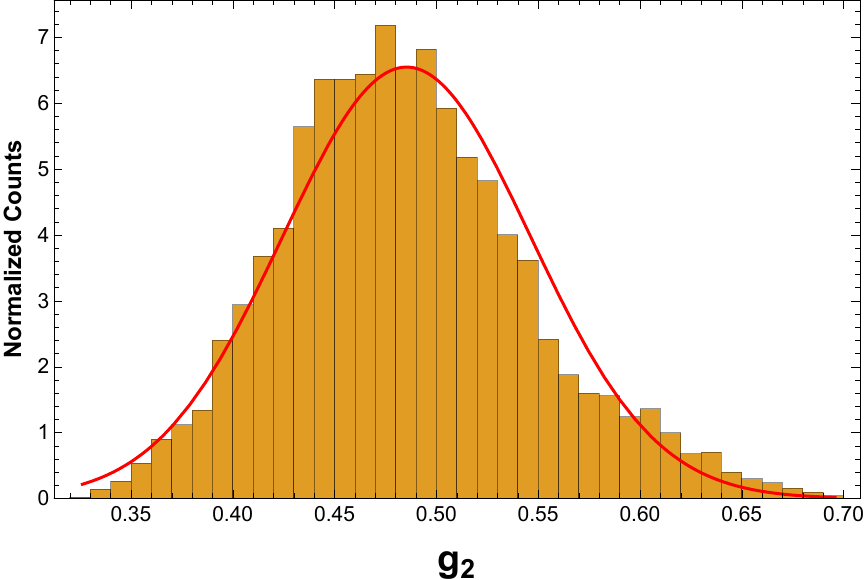}
\includegraphics[width=0.32\textwidth]{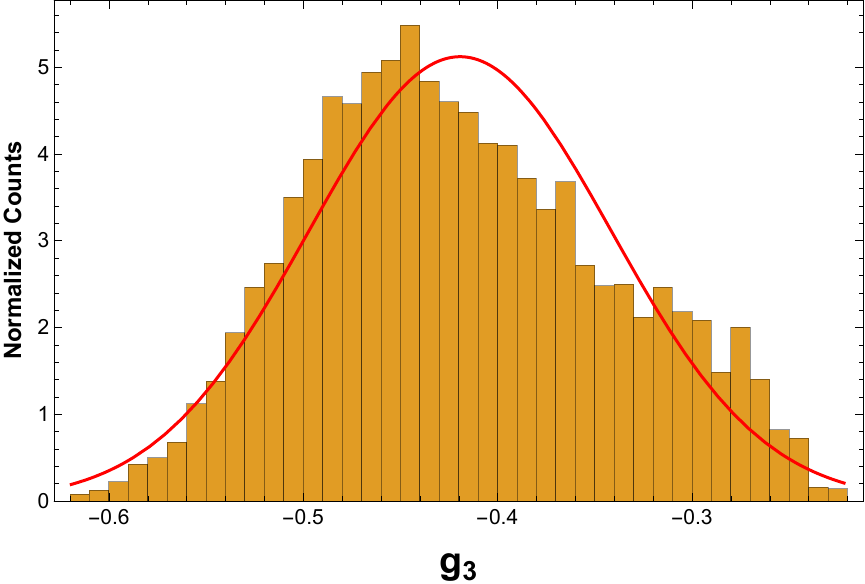}
\caption{Normalized distributions of the $\Xi_c^0 \to \Xi^- \ell^+ \nu$ 
form factors $f_i$ and $g_i$ at $q^2 = 0$ obtained from LCSR. The solid lines 
represent Gaussian fits to the Monte Carlo distributions.}
\label{fig:xic0_to_ximn_form}
\end{figure}
From this comparison, we make the following observations:
Our prediction for the branching ratio of $\Xi_c^0 \to \Xi^- e^+ \nu_e$ shows 
excellent agreement with the recent lattice QCD result~\cite{Farrell:2025gis}, 
but deviates from other theoretical predictions—including LCSR results based on 
the distribution amplitudes of $\Xi$ baryons~\cite{Azizi:2011mw,Aliev:2021wat}—as 
well as from current experimental measurements.
 
These findings underscore a possible tension between theoretical predictions and 
experimental data for the $\Xi_c \to \Xi \ell \nu$ channels. This discrepancy 
calls for both improved experimental measurements and refined theoretical 
approaches. There are two possible sources that could help clarify 
the situation:

\begin{enumerate}[label=\alph*)]
\item As previously noted, the branching ratio $\Xi_c \to \Xi \ell \nu$ is often 
inferred using experimental data for the nonleptonic decay $\Xi_c \to \Xi \pi$. 
Recent SU(3) flavor symmetry analyses of charm decays suggest that the measured 
value of $\mathcal{B}(\Xi_c^0 \to \Xi^- \pi^+)$ may be underestimated~\cite{Geng:2023pkr,Xing:2024nvg}. 
Therefore, a new precise measurement of this decay mode is essential for resolving the tension. 
\item A more precise determination of the distribution amplitudes (DAs) would significantly 
enhance the reliability of QCD-based predictions, including those obtained via light-cone sum rules. 
\end{enumerate}

\begin{table}[htbp]
\centering
\small
\caption{Existing experimental and theoretical results for the branching ratios (in \%) of 
the semileptonic $\Xi_c \to \Xi \ell \nu$ decays.}
\begin{tabular}{lcccc}
\toprule
\textbf{} & 
$\Xi_c^0 \to \Xi^- e^+ \nu_e$ & 
$\Xi_c^0 \to \Xi^- \mu^+ \nu_\mu$ & 
$\Xi_c^+ \to \Xi^0 e^+ \nu_e$ & 
$\Xi_c^+ \to \Xi^0 \mu^+ \nu_\mu$  \\  
\midrule
\textbf{This Work} 
& $3.73 \pm 1.04 $ & $3.59 \pm 1.01$ & $11.2 \pm 3.25$ & $10.8 \pm 3.13$   \\
LCSR~\cite{Aliev:2021wat} 
& $1.85 \pm 0.56$ & $1.79 \pm 0.54$ & $5.51 \pm 1.65$ & $5.34 \pm 1.61$  \\
BELLE II~\cite{Belle:2021crz} 
& $1.72 \pm 0.10 \pm 0.12 \pm 0.50$ & $1.71 \pm 0.17 \pm 0.13 \pm 0.50$ & --- & ---  \\
ALICE~\cite{ALICE:2021bli} 
& $1.8 \pm 0.2$ & $1.8 \pm 0.2$ & --- & ---  \\
SU(3)~\cite{Geng:2018plk} 
& $4.87 \pm 1.74$ & --- & $3.38^{+2.10}_{-2.26}$ & ---  \\
SU(3)~\cite{Geng:2019bfz} 
& $2.4 \pm 0.3$ & $2.4 \pm 0.3$ & $9.8 \pm 1.1$ & $9.8 \pm 1.1$  \\
RQM~\cite{Faustov:2019ddj} 
& 2.38 & 2.31 & 9.40 & 9.11  \\
LATTICE~\cite{Zhang:2021oja} 
& $2.38 \pm 0.33$ & $2.29 \pm 0.29 \pm 0.33$ & $7.18 \pm 0.90 \pm 0.98$ & $6.91 \pm 0.87 \pm 0.93$  \\
LATTICE~\cite{Farrell:2025gis} 
& $3.58 \pm 0.12$ & $3.47 \pm 0.12 $ & $10.94 \pm 0.34$ & $10.61 \pm 0.33$  \\
3PSR~\cite{Zhao:2021sje} 
& $1.45 \pm 0.31$ & $1.45 \pm 0.31$ & --- & ---  \\
LCSR~\cite{Azizi:2011mw} 
& $7.26 \pm 2.54$ & $7.15 \pm 2.50$ & $28.6 \pm 10$ & $28.2 \pm 9.9$  \\
LFQM~\cite{Zhao:2018zcb} 
& 1.354 & --- & 5.39 & ---  \\
LF~\cite{Ke:2021pxk} 
& $1.72 \pm 0.35$ & --- & $5.2 \pm 1.02$ & ---  \\
\bottomrule
\end{tabular}
\label{tab:comparison}
\end{table}

\section{Conclusion}
\label{sec:conclusion}

In this work, we presented a new light-cone QCD sum rule (LCSR) analysis of the 
semileptonic decays $\Xi_c \to \Xi \ell^+ \nu_\ell$, using the distribution amplitudes (DAs) of 
the initial $\Xi_c$ baryon as the primary nonperturbative input. Our predictions 
for the branching ratios—particularly for the neutral channel 
$\Xi_c^0 \to \Xi^- \ell^+ \nu$—are notably larger than current experimental 
measurements, yet they are in good agreement with recent lattice QCD results. 
This consistency suggests a possible tension between theoretical predictions 
and experimental extractions, which often rely on indirect determinations via 
normalization to hadronic decay modes. A remeasurement of 
$\mathcal{B}(\Xi_c^0 \to \Xi^- \pi^+)$ or a direct measurement of 
$\mathcal{B}(\Xi_c^0 \to \Xi^- \ell^+ \nu)$ would help resolve this ambiguity, 
while a more precise determination of the distribution amplitudes of heavy baryons 
would provide a firmer benchmark for theoretical frameworks. Overall, our results 
show that further experimental and theoretical investigations are necessary to 
improve our understanding of semileptonic charmed baryon decays.


\newpage

\bibliographystyle{utcaps_mod}
\bibliography{all.bib}

\providecommand{\href}[2]{#2}\begingroup\raggedright\begin{thebibliography}{10}

\bibitem{Belle:2018kzz}
{\bfseries Belle} Collaboration, Y.~B. Li {\em et~al.}, ``{\em {First
  Measurements of Absolute Branching Fractions of the $\Xi_c^0$ Baryon at
  Belle}},'' \href{http://dx.doi.org/10.1103/PhysRevLett.122.082001}{Phys. Rev.
  Lett. {\bfseries 122} no.~8, (2019) 082001},
  \href{http://arxiv.org/abs/1811.09738}{[{\ttfamily 1811.09738}]}.

\bibitem{Belle:2021crz}
{\bfseries Belle} Collaboration, Y.~B. Li {\em et~al.}, ``{\em {Measurements of
  the branching fractions of the semileptonic decays $\Xi_{c}^{0} \to \Xi^{-}
  \ell^{+} \nu_{\ell}$ and the asymmetry parameter of $\Xi_{c}^{0} \to \Xi^{-}
  \pi^{+}$}},'' \href{http://dx.doi.org/10.1103/PhysRevLett.127.121803}{Phys.
  Rev. Lett. {\bfseries 127} no.~12, (2021) 121803},
  \href{http://arxiv.org/abs/2103.06496}{[{\ttfamily 2103.06496}]}.

\bibitem{ALICE:2021bli}
{\bfseries ALICE} Collaboration, S.~Acharya {\em et~al.}, ``{\em {Measurement
  of the Cross Sections of $\Xi^0_{c}$ and $\Xi^+_{c}$ Baryons and of the
  Branching-Fraction Ratio BR($\Xi^0_{c} \rightarrow \Xi^-{e}^+\nu_{
  e}$)/BR($\Xi^0_{c} \rightarrow \Xi^-\pi^+$) in pp collisions at 13 TeV}},''
  \href{http://dx.doi.org/10.1103/PhysRevLett.127.272001}{Phys. Rev. Lett.
  {\bfseries 127} no.~27, (2021) 272001},
  \href{http://arxiv.org/abs/2105.05187}{[{\ttfamily 2105.05187}]}. [Erratum:
  Phys.Rev.Lett. 134, 179902 (2025)].

\bibitem{Farrell:2025gis}
C.~Farrell and S.~Meinel, ``{\em {$\Xi_c \to \Xi$ form factors from lattice QCD
  with domain-wall quarks: A new piece in the puzzle of $\Xi_c^0$ decay
  rates}},'' \href{http://dx.doi.org/10.1103/35c9-kbvr}{Phys. Rev. D {\bfseries
  111} no.~11, (2025) 114521},
  \href{http://arxiv.org/abs/2504.07302}{[{\ttfamily 2504.07302}]}.

\bibitem{ParticleDataGroup:2020ssz}
{\bfseries Particle Data Group} Collaboration, P.~A. Zyla {\em et~al.}, ``{\em
  {Review of Particle Physics}},''
  \href{http://dx.doi.org/10.1093/ptep/ptaa104}{PTEP {\bfseries 2020} no.~8,
  (2020) 083C01}.

\bibitem{Ebert:2011kk}
D.~Ebert, R.~N. Faustov, and V.~O. Galkin, ``{\em {Spectroscopy and Regge
  trajectories of heavy baryons in the relativistic quark-diquark picture}},''
  \href{http://dx.doi.org/10.1103/PhysRevD.84.014025}{Phys. Rev. D {\bfseries
  84} (2011) 014025}, \href{http://arxiv.org/abs/1105.0583}{[{\ttfamily
  1105.0583}]}.

\bibitem{He:2021qnc}
X.-G. He, F.~Huang, W.~Wang, and Z.-P. Xing, ``{\em {SU(3) symmetry and its
  breaking effects in semileptonic heavy baryon decays}},''
  \href{http://dx.doi.org/10.1016/j.physletb.2021.136765}{Phys. Lett. B
  {\bfseries 823} (2021) 136765},
  \href{http://arxiv.org/abs/2110.04179}{[{\ttfamily 2110.04179}]}.

\bibitem{Geng:2019bfz}
C.-Q. Geng, C.-W. Liu, T.-H. Tsai, and S.-W. Yeh, ``{\em {Semileptonic decays
  of anti-triplet charmed baryons}},''
  \href{http://dx.doi.org/10.1016/j.physletb.2019.03.056}{Phys. Lett. B
  {\bfseries 792} (2019) 214--218},
  \href{http://arxiv.org/abs/1901.05610}{[{\ttfamily 1901.05610}]}.

\bibitem{Geng:2017mxn}
C.~Q. Geng, Y.~K. Hsiao, C.-W. Liu, and T.-H. Tsai, ``{\em {Charmed Baryon Weak
  Decays with SU(3) Flavor Symmetry}},''
  \href{http://dx.doi.org/10.1007/JHEP11(2017)147}{JHEP {\bfseries 11} (2017)
  147}, \href{http://arxiv.org/abs/1709.00808}{[{\ttfamily 1709.00808}]}.

\bibitem{Geng:2018plk}
C.~Q. Geng, Y.~K. Hsiao, C.-W. Liu, and T.-H. Tsai, ``{\em {Antitriplet charmed
  baryon decays with SU(3) flavor symmetry}},''
  \href{http://dx.doi.org/10.1103/PhysRevD.97.073006}{Phys. Rev. D {\bfseries
  97} no.~7, (2018) 073006}, \href{http://arxiv.org/abs/1801.03276}{[{\ttfamily
  1801.03276}]}.

\bibitem{Faustov:2018ahb}
R.~N. Faustov and V.~O. Galkin, ``{\em {Relativistic description of the $\Xi_b$
  baryon semileptonic decays}},''
  \href{http://dx.doi.org/10.1103/PhysRevD.98.093006}{Phys. Rev. D {\bfseries
  98} no.~9, (2018) 093006}, \href{http://arxiv.org/abs/1810.03388}{[{\ttfamily
  1810.03388}]}.

\bibitem{Perez-Marcial:1989sch}
R.~Perez-Marcial, R.~Huerta, A.~Garcia, and M.~Avila-Aoki, ``{\em {Predictions
  for Semileptonic Decays of Charm Baryons. 2. Nonrelativistic and {MIT} Bag
  Quark Models}},'' \href{http://dx.doi.org/10.1103/PhysRevD.44.2203}{Phys.
  Rev. D {\bfseries 40} (1989) 2955}. [Erratum: Phys.Rev.D 44, 2203 (1991)].

\bibitem{Ivanov:1996fj}
M.~A. Ivanov, V.~E. Lyubovitskij, J.~G. Korner, and P.~Kroll, ``{\em {Heavy
  baryon transitions in a relativistic three quark model}},''
  \href{http://dx.doi.org/10.1103/PhysRevD.56.348}{Phys. Rev. D {\bfseries 56}
  (1997) 348--364}, \href{http://arxiv.org/abs/hep-ph/9612463}{[{\ttfamily
  hep-ph/9612463}]}.

\bibitem{Zhang:2021oja}
Q.-A. Zhang {\em et~al.}, ``{\em {First lattice QCD calculation of semileptonic
  decays of charmed-strange baryons {\ensuremath{\Xi}}$_{c}$ *}},''
  \href{http://dx.doi.org/10.1088/1674-1137/ac2b12}{Chin. Phys. C {\bfseries
  46} no.~1, (2022) 011002}, \href{http://arxiv.org/abs/2103.07064}{[{\ttfamily
  2103.07064}]}.

\bibitem{Briceno:2012wt}
R.~A. Briceno, H.-W. Lin, and D.~R. Bolton, ``{\em {Charmed-Baryon Spectroscopy
  from Lattice QCD with $N_f$ = 2+1+1 Flavors}},''
  \href{http://dx.doi.org/10.1103/PhysRevD.86.094504}{Phys. Rev. D {\bfseries
  86} (2012) 094504}, \href{http://arxiv.org/abs/1207.3536}{[{\ttfamily
  1207.3536}]}.

\bibitem{Zhao:2021sje}
Z.-X. Zhao, X.-Y. Sun, F.-W. Zhang, Y.-P. Xing, and Y.-T. Yang, ``{\em
  {Semileptonic form factors of
  {\ensuremath{\Xi}}c{\textrightarrow}{\ensuremath{\Xi}} in QCD sum rules}},''
  \href{http://dx.doi.org/10.1103/PhysRevD.108.116008}{Phys. Rev. D {\bfseries
  108} no.~11, (2023) 116008},
  \href{http://arxiv.org/abs/2103.09436}{[{\ttfamily 2103.09436}]}.

\bibitem{Liu:2010bh}
Y.-L. Liu and M.-Q. Huang, ``{\em {A light-cone QCD sum rule approach for the
  $\Xi$ baryon electromagnetic form factors and the semileptonic decay $\Xi_c
  \to \Xi e^+ \nu_e$ }},''
  \href{http://dx.doi.org/10.1088/0954-3899/37/11/115010}{J. Phys. G {\bfseries
  37} (2010) 115010}, \href{http://arxiv.org/abs/1102.4245}{[{\ttfamily
  1102.4245}]}.

\bibitem{Azizi:2011mw}
K.~Azizi, Y.~Sarac, and H.~Sundu, ``{\em {Light cone QCD sum rules study of the
  semileptonic heavy $\Xi_{Q}$ and $\Xi'_{Q}$ transitions to $\Xi$ and $\Sigma
  $ baryons}},'' \href{http://dx.doi.org/10.1140/epja/i2012-12002-1}{Eur. Phys.
  J. A {\bfseries 48} (2012) 2},
  \href{http://arxiv.org/abs/1107.5925}{[{\ttfamily 1107.5925}]}.

\bibitem{Aliev:2021wat}
T.~M. Aliev, S.~Bilmis, and M.~Savci, ``{\em {Semileptonic {\ensuremath{\Xi}}c
  baryon decays in the light cone QCD sum rules}},''
  \href{http://dx.doi.org/10.1103/PhysRevD.104.054030}{Phys. Rev. D {\bfseries
  104} no.~5, (2021) 054030},
  \href{http://arxiv.org/abs/2108.01378}{[{\ttfamily 2108.01378}]}.

\bibitem{Shifman:1978bx}
M.~A. Shifman, A.~I. Vainshtein, and V.~I. Zakharov, ``{\em {QCD and Resonance
  Physics. Theoretical Foundations}},''
  \href{http://dx.doi.org/10.1016/0550-3213(79)90022-1}{Nucl. Phys. B
  {\bfseries 147} (1979) 385--447}.

\bibitem{Shifman:1978by}
M.~A. Shifman, A.~I. Vainshtein, and V.~I. Zakharov, ``{\em {QCD and Resonance
  Physics: Applications}},''
  \href{http://dx.doi.org/10.1016/0550-3213(79)90023-3}{Nucl. Phys. B
  {\bfseries 147} (1979) 448--518}.

\bibitem{Chung:1981cc}
Y.~Chung, H.~G. Dosch, M.~Kremer, and D.~Schall, ``{\em {Baryon Sum Rules and
  Chiral Symmetry Breaking}},''
  \href{http://dx.doi.org/10.1016/0550-3213(82)90154-7}{Nucl. Phys. B
  {\bfseries 197} (1982) 55--75}.

\bibitem{Ali:2012pn}
A.~Ali, C.~Hambrock, A.~Y. Parkhomenko, and W.~Wang, ``{\em {Light-Cone
  Distribution Amplitudes of the Ground State Bottom Baryons in HQET}},''
  \href{http://dx.doi.org/10.1140/epjc/s10052-013-2302-4}{Eur. Phys. J. C
  {\bfseries 73} no.~2, (2013) 2302},
  \href{http://arxiv.org/abs/1212.3280}{[{\ttfamily 1212.3280}]}.

\bibitem{Gubernari:2018wyi}
N.~Gubernari, A.~Kokulu, and D.~van Dyk, ``{\em {$B\to P$ and $B\to V$ Form
  Factors from $B$-Meson Light-Cone Sum Rules beyond Leading Twist}},''
  \href{http://dx.doi.org/10.1007/JHEP01(2019)150}{JHEP {\bfseries 01} (2019)
  150},
\href{http://arxiv.org/abs/1811.00983}{[{\ttfamily 1811.00983}]}.

\bibitem{Aliev:2019ojc}
T.~M. Aliev, H.~Dag, A.~Kokulu, and A.~Ozpineci, ``{\em {$B \to T$ transition
  form factors in light-cone sum rules}},''
  \href{http://dx.doi.org/10.1103/PhysRevD.100.094005}{Phys. Rev. {\bfseries
  D100} no.~9, (2019) 094005},
\href{http://arxiv.org/abs/1908.00847}{[{\ttfamily 1908.00847}]}.

\bibitem{Gutsche:2015mxa}
T.~Gutsche, M.~A. Ivanov, J.~G. K{\"o}rner, V.~E. Lyubovitskij, P.~Santorelli,
  and N.~Habyl, ``{\em {Semileptonic decay $\Lambda_b \to \Lambda_c + \tau^- +
  \bar{\nu_\tau}$ in the covariant confined quark model}},''
  \href{http://dx.doi.org/10.1103/PhysRevD.91.074001}{Phys. Rev. D {\bfseries
  91} no.~7, (2015) 074001}, \href{http://arxiv.org/abs/1502.04864}{[{\ttfamily
  1502.04864}]}. [Erratum: Phys.Rev.D 91, 119907 (2015)].

\bibitem{Bourrely:2008za}
C.~Bourrely, I.~Caprini, and L.~Lellouch, ``{\em {Model-independent description
  of $B \to \pi l \nu$ decays and a determination of $|V_{ub}|$}},''
  \href{http://dx.doi.org/10.1103/PhysRevD.82.099902}{Phys. Rev. D {\bfseries
  79} (2009) 013008}, \href{http://arxiv.org/abs/0807.2722}{[{\ttfamily
  0807.2722}]}. [Erratum: Phys.Rev.D 82, 099902 (2010)].

\bibitem{Geng:2023pkr}
C.-Q. Geng, X.-G. He, X.-N. Jin, C.-W. Liu, and C.~Yang, ``{\em {Complete
  determination of SU(3)F amplitudes and strong phase in $\Lambda_c^+ \to \Xi_0
  K^+$}},'' \href{http://dx.doi.org/10.1103/PhysRevD.109.L071302}{Phys. Rev. D
  {\bfseries 109} no.~7, (2024) L071302},
  \href{http://arxiv.org/abs/2310.05491}{[{\ttfamily 2310.05491}]}.

\bibitem{Xing:2024nvg}
Z.-P. Xing, Y.-J. Shi, J.~Sun, and Y.~Xing, ``{\em {SU(3) symmetry analysis in
  charmed baryon two body decays with penguin diagram contribution}},''
  \href{http://dx.doi.org/10.1140/epjc/s10052-024-13389-y}{Eur. Phys. J. C
  {\bfseries 84} no.~10, (2024) 1014},
  \href{http://arxiv.org/abs/2407.09234}{[{\ttfamily 2407.09234}]}.

\bibitem{Faustov:2019ddj}
R.~N. Faustov and V.~O. Galkin, ``{\em {Semileptonic $\Xi_c$ baryon decays in
  the relativistic quark model}},''
  \href{http://dx.doi.org/10.1140/epjc/s10052-019-7214-5}{Eur. Phys. J. C
  {\bfseries 79} no.~8, (2019) 695},
  \href{http://arxiv.org/abs/1905.08652}{[{\ttfamily 1905.08652}]}.

\bibitem{Zhao:2018zcb}
Z.-X. Zhao, ``{\em {Weak decays of heavy baryons in the light-front
  approach}},'' \href{http://dx.doi.org/10.1088/1674-1137/42/9/093101}{Chin.
  Phys. C {\bfseries 42} no.~9, (2018) 093101},
  \href{http://arxiv.org/abs/1803.02292}{[{\ttfamily 1803.02292}]}.

\bibitem{Ke:2021pxk}
H.-W. Ke, Q.-Q. Kang, X.-H. Liu, and X.-Q. Li, ``{\em {Weak decays of in the
  light-front quark model *}},''
  \href{http://dx.doi.org/10.1088/1674-1137/ac1c66}{Chin. Phys. C {\bfseries
  45} no.~11, (2021) 113103},
  \href{http://arxiv.org/abs/2106.07013}{[{\ttfamily 2106.07013}]}.

\end{thebibliography}\endgroup


\end{document}